\documentclass[12pt]{spieman}
\usepackage[utf8]{inputenc}
\usepackage{authblk}
\usepackage[square,sort,comma,numbers]{natbib}
\usepackage{graphicx}
\usepackage{float}
\usepackage{gensymb}
\usepackage{url}
\usepackage{caption}
\usepackage{subcaption}
\usepackage{amssymb}
\usepackage{amsmath}
\usepackage{bm}
\usepackage{siunitx}
\usepackage[normalem]{ulem}
\usepackage{wrapfig}
\usepackage{color}
\usepackage{upgreek}
\usepackage{booktabs}
\usepackage{lineno}
\usepackage{multirow}
\usepackage{pbox}
\usepackage{geometry}
\title{Optimum telescope focal ratios for microlens-to-fiber coupled integral field spectrographs}

\author[a,*]{Sabyasachi Chattopadhyay}
\author[a,b,c]{Matthew A. Bershady}
\author[b]{Marsha J. Wolf}
\author[b]{Michael P. Smith}

\affil[a]{South African Astronomical Observatory, 1 Observatory Rd, Observatory, Cape Town, 7925, South Africa}

\affil[b]{University of Wisconsin, Department of Astronomy, 475 North Charter Street, Madison, WI 53706, USA}

\affil[c]{Department of Astronomy, University of Cape Town, Private Bag X3, Rondebosch 7701, South Africa}

\begin{document} 

\maketitle

\begin{abstract}

We describe the optimum telescope focal ratio for a two-element, three surface, telecentric image-transfer microlens-to-fiber coupled integral field unit within the constraints imposed by micro-optics fabrication and optical aberrations. We create a generalized analytical description of the micro-optics optical parameters from first principals. We find that the optical performance, including all aberrations, of a design constrained by an analytic model considering only spherical aberration and diffraction matches within $\pm$4\% of a design optimized by ray-tracing software such as Zemax. The analytical model does not require any compromise on the available clear aperture; about 90\% mechanical aperture of a hexagonal microlens is available for light collection. The optimum telescope f-ratio for a 200$\upmu$m core fiber fed at f/3.5 is between f/7 and f/12. We find the optimum telescope focal ratio changes as a function of fiber core diameter and fiber input beam speed. A telescope focal ratio of f/8 would support the largest range of fiber diameters (100 to 500$\upmu$m) and fiber injection speeds (between f/3 and f/5). The optimization of telescope and lenslet-coupled fibers is relevant for the design of high-efficiency dedicated survey telescopes, and for retro-fitting existing facilities via introducing focal macro-optics to match the instrument input requirements.

\end{abstract}

\keywords{Integral Field Spectroscopy; Microlens-Fiber IFU; Microlens Optical Design: Telescope f-ratio.}

{\noindent \footnotesize \textbf{*} {Further author information: (Send correspondence to Sabysachi Chattopadhyay)\\E-mail: sabyasachi@saao.ac.za, Telephone: +27 78 298 6850} }
%\begin{spacing}{2}
\section{Introduction}
\label{sec:intro}

Modern day spectroscopic surveys employ collecting apertures running from   10m-class telescopes (VIRUS\cite{virusp,virusw}, PFS\cite{pfs}, MSE\cite{MSE}), to modest 4m-class telescopes (e.g. DESI\cite{desi}, 4MOST\cite{4most}), down to even a few hundred mm (LVM\cite{lvm}). At the heart of each of these surveys is fiber-optic coupled spectroscopy. Fiber optics serve as a convenient bridge between the mobile telescope system and the more stable spectrograph mounts, as well as a image reformatting system from the telescope focal plane to spectrograph entrance slit. While Galactic and extragalactic surveys have traditionally employed single fibers to multiplex redshift and spectral diagnostic measurements of stars, distant galaxies and quasars, more recently there has been an additional focus on mapping resolved sources (nearby galaxies and Galactic nebulae) using integral-field spectroscopy (IFS). In general, fiber-fed IFS (IFUs; e.g., SparsePak\cite{bershady}, PPak\cite{kelz}, VIRUS-P\cite{virusp}, VIRUS-W\cite{virusw}, MaNGA\cite{drory}, and MEGARA\cite{megara}) are flexible, cost-effective and suitable for optical and near IR wavelengths. Fibers also lend themselves well to multi-object IFS of distributed and/or extended sources contained in large (often many deg$^2$) telescope fields of view (e.g., SAMI\cite{SAMI} on the 4m AAT and MaNGA\cite{manga} on the 2.5m Sloan Telescope).  Consequently fiber-fed IFS remains an appealing approach along side other flavours of IFS (e.g., image-slicing with SPHERE\cite{sphere}, MUSE\cite{vlt} and KMOS\cite{kmos} on VLT, or KCWI\cite{cwi} on Keck).

Despite fiber optics being a popular and cost-effective mode of coupling telescopes to spectrographs, fibers are well-known to degrade the input focal-ratio upon output. This focal ratio degradation (FRD) via stepped index, multi-mode fibers\cite{arthur,alsmith,carasco} affects the efficiency of observation because it is in essence an injection of entropy into the information gathering system. It is observed \cite{parry,Ramsey1988} that the effect of FRD can be minimized by injecting a faster beam into the fiber. 

\begin{figure}[h]
\centering
\includegraphics[width = \linewidth]{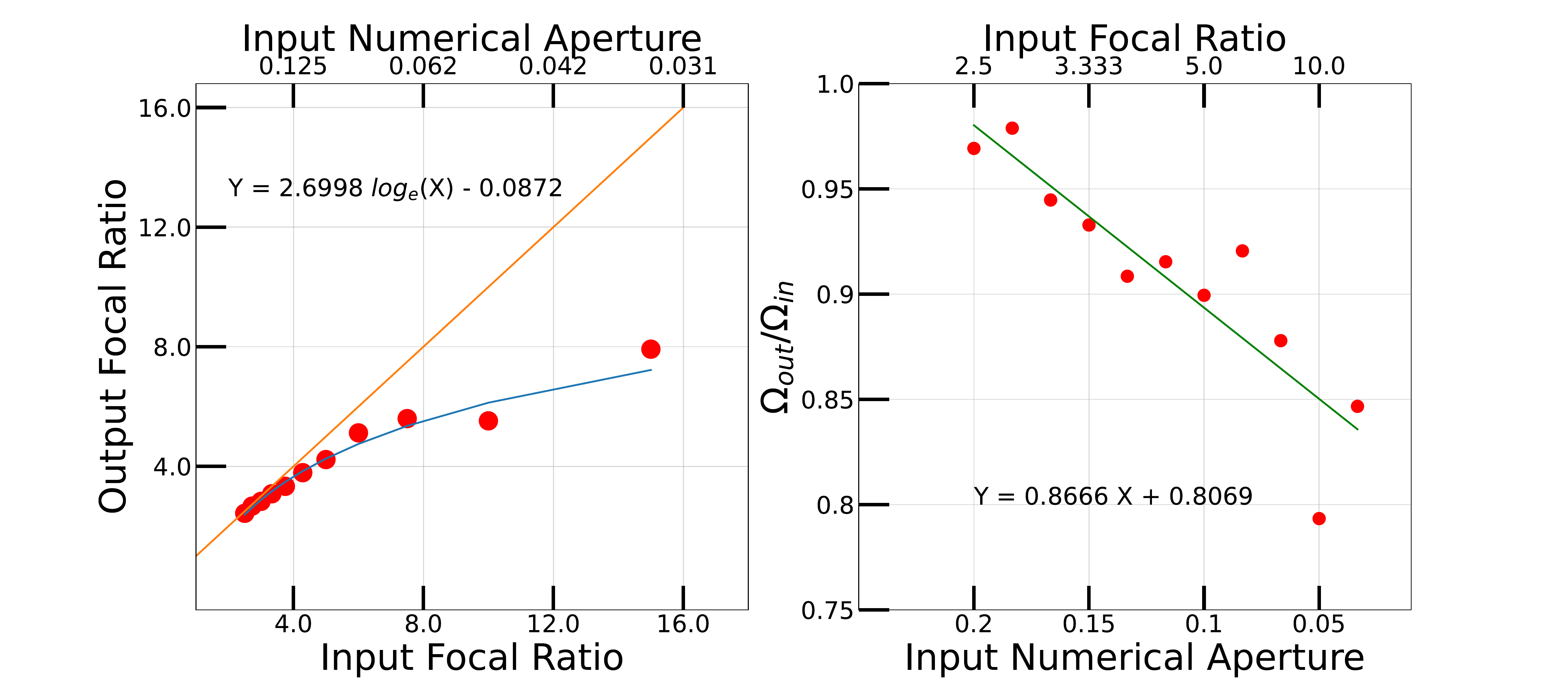}
\caption{Focal ratio degradation by a 2m long 400$\upmu$m diameter core fiber for various input focal ratio measured at a normalized encircled energy of 98\%. Measurements details are described in the text. The figure depicts the effect of focal ratio degradation via fiber increases with slower input beam in the left panel while the right panel shows the linear loss of solid angle on sky with decreasing input numerical aperture.}
\label{fig:frd}
\end{figure}

A contemporary measurement of this effect is illustrated in Figure \ref{fig:frd} using a differential reimaging system similar to  \cite{bershady,drory,murphy}. The focal ratio degradation was measured for a 2-m length of 400$\upmu$m diameter core step-index, broad-band fiber (MOLEX/Polymicro FBP 400:440:470) fed with a centered, reimaged near-field spot of 100$\upmu$m in diameter. The input focal ratio was varied by changing the aperture of an intermediate pupil-stop. The output focal ratio was measured at a radius covering 98\% of the normalized encircled energy. The salient feature of this measurement was that it was performed with a well-aligned inject beam, telecentric to within 0.1$\degree$ in tip/tilt similar to that of \cite{murphy}. The upper limit on the angular deviation limits geometric FRD (gFRD) to no faster than f/286; over the range of injection speeds this is inconsequential. While the left hand panel of Figure \ref{fig:frd} shows the well-known non-linear `saturation' of the output beam speed, in the right panel we present the same measurements as a ratio of out vs input solid angle. This has a simple, linear behavior with input numerical aperture from which entropy gain due to fiber FRD can be easily estimated. The qualitative behavior motivates why present-day fiber-coupled spectrographs tend to have faster collimators (see the references above).

It is also the case that for IFS using bare, densely packed circular fibers, the fill factor is typically only 60\% \cite{drory} given the inherent geometry plus the minimum requirements for cladding and mechanical buffer. Some gains can be had by removing the buffer and lightly fusing the fibers, but is technically challenging and comes at the cost of more FRD \cite{Bryant14}. Consequently, it is often desirable to couple fibers to the telescope focal plane via lenslet arrays. One purpose of lenslet coupling is to increase the fill-factor of the fiber array.  In principle, circular lenslets with hexagonal packing or hexagonal lenslets with circular chromium coating can increase the fill-factor up to 90\%. We consider this lenslet geometry in this paper. However, when coupling lenslets to fibers, a further challenge arises: The clear aperture in a micro-lens system falls quickly compared to the available mechanical aperture with faster input beams from telescope \cite{sabyasachi2021, chattopadhyay2020}. Thus in an effort to reduce FRD one would struggle to attain an acceptable fill factor on sky \cite{sabyasachi2021}.

The aim of this study is to resolve the tension between the optimum input focal ratios for lenslet and fiber optics. This is part of the broader question i.e. given a survey being planned, either using an existing telescope or a new telescope, what would be the ideal telescope focal ratio at the input of the microlens for a set of fiber input speed, fiber diameter and fabricable microlens properties. To do so we develop a generic analytical description of an image transfer IFU optical model to optimize the telescope focal ratio within the constraints posed by the microlens array fabrication. We assume the spectrograph design accommodates a fast fiber output beam or, alternatively, uses additional microlenses at the output end of the fiber to modulate the speed of the spectrograph collimator. We start in Section 2 by describing the optical parameters of an image transfer IFU system and the parameter space available for optimization given fabrication limitations and FRD. In Section 3 we describe the analytical model to define the IFU optical parameters. Following sections discuss two distinct regimes where on-axis aberrations are either dominated by diffraction or spherical aberration (SA) (Section 4) and the design choices that minimize light loss from aberrations in general (Section 5). We confirm the reliability of our model compared to ray-trace calculations in the non-diffraction-limited regime in Section 5 as well. In Section 6 we describe the acceptable range of telescope beam speed defined by the fabrication limits. Section 7 summarizes our results and presents our conclusions.

The result of our analysis calls for much slower telescope beams than developed for telescopes optimized for bare-fiber feeds, such as HET and SALT, which are close to f/4. The optimum telescope f-ratio we find is closer to f/8, although in detail we will show this depends on the fiber core diameter. The optimum f-ratio can be achieved on an existing telescope via a macro-focal reducer or expander \cite{HLee2018, HLee2020, GHill2020}. However, for a dedicated survey facility we suggest optimizing the native telescope beam speed to minimize the total number of optical elements \cite{Chonis2012, HLee2010}. 

\section{Design choices and constraints} 
\label{sec:choices}

\subsection{Image transfer microlens system}
\label{sec:MLAsys}

We describe telecentric reimaging (or image-relay) microlens systems in this paper, and note differences with pupil-transfer microlens systems below. The basic optical concept is to re-image the focal plane on to the fiber cores through a set of two MLAs. Conceptually, the first MLA elements act as field lenses while the second produces telecentric images. In practice, to optimize image quality the optics deviate slightly from this idealized prescription [see for example VIRUS2 design and Section~\ref{sec:minloss} here]\cite{Hill2020}. A bi-convex MLA (BC) with appropriate radius of curvature (RoC) and thickness creates a pupil at its exit surface, and the exit beam is collimated for all field points. A suitable gap (depending on the RoC of the second MLA) of the BC and plano-convex MLA (PC) ensures a telecentric image at the flat back surface of the PC. 

One of the free design parameter is the micro-image size at the input of the fiber. However, in our previous study \cite{chattopadhyay2020}, we found that the micro-image diameter should be $\sim$97\% of the fiber core diameter in order to minimize \'etendue gain. Thus the fiber core diameter and micro-image diameter are used interchangably in our analysis. To maximize grasp (area--solid-angle product, or A$\Omega$), the ideal micro-image diameter ($\rm d_m$) should be equal to the diameter of the fiber core. However, due to fiber positioning and alignment issues $\rm d_m$ needs to be smaller than fiber core. Based on our own photo-lithographic work and comparison to commercially available products we estimate current technology permits fibers to be positioned within $\rm \pm 3~\upmu$m RMS at any desired spacing\cite{chattopadhyay2020}. 

\begin{figure}[h]
\centering
\includegraphics[width = 0.9\linewidth]{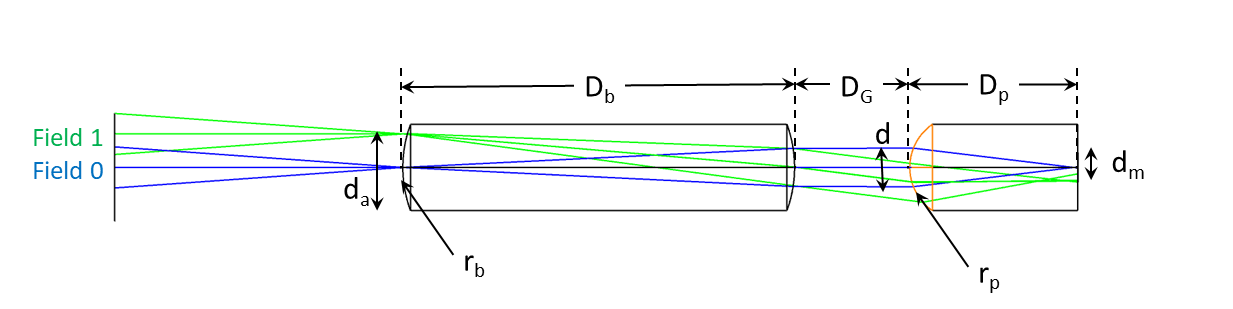}
\caption{Arrangement of a reimaging microlens array system that captures a part of the telescope focal plane ($\rm d_a$) and injects this into the fiber core. Key design parameters are: (a) Field 0: Central field point; (b) Field 1: Edge field point; (c) $\rm{d_a}$: Diameter of the input beam/Bi-convex microlens (BC) clear aperture diameter; (d) $\rm {r_b}$: BC radius of curvature; (e) $\rm{D_b}$: BC thickness; (f) d: Field 0 pupil diameter; (g)  $\rm {r_p}$: plano-convex microlens (PC) radius of curvature; (h) $\rm{D_p}$: PC thickness; (i) $\rm{D_G}$: Air gap width between PC and BC; (j) $\rm{d_m}$: micro-image diameter/fiber core diameter. Details about implemented microlens properties are discussed in \cite{Lee_2001}.}
\label{fig:mla}
\end{figure}

Figure~\ref{fig:mla} illustrates a typical image-relay microlens system where telescope light is fed at a focal ratio of $\rm f_{\rm tel}$. This beam enters the bi-convex microlens with $\rm r_b$ radius of curvature, $\rm D_b$ thickness, and $\rm d_a$ clear aperture. The thickness is defined such that the beam forms its pupil at the exit surface of the bi-convex microlens. The pupil is then focused to an image plane via a plano-convex lens with $\rm r_p$ radius of curvature and $\rm D_p$ thickness, sitting at a distance of $\rm D_G$ from the bi-convex lens. $\rm D_p$ and the beam diameter of the central field defines the injected f-ratio ($\rm f_{\rm fib}$) into the fiber. Since the pupil is positioned at the exit surface of the bi-convex MLA, one can define $\rm D_p$ as $\rm n_g \times r_p$/($\rm n_g$ - 1) so that the plano-convex microlens produce telecentric micro-image at its planar back plane.  Thus $\rm f_{\rm fib}$ and hence $\rm d$ and $\rm D_p$ define $\rm r_p$ as well. On the other hand, $\rm D_G$ defines the MLA clear aperture diameter which in turn defines the sky sampling size (via plate scale). We provide mathematical prescriptions of all these dependencies in the following section \ref{sec:params} and more details about their derivation can be found in Appendix \ref{app:derivation}.

\subsection{Other microlens systems}
\label{sec:otherMLA}

Other microlens relay options include transfer of the image with a pupil onto the fiber entrance aperture. These so-called pupil-transfer systems have the advantage of requiring only one MLA. In the simplest case using a single plano-convex MLA can be bonded directly to the fiber to minimize reflection losses. Unfortunately, the pupil image onto the fiber is not telecentric, leading to potentially significant gFRD. Because the fibers azimuthally scramble the input signal, the non-telecentric feed cannot be corrected at the fiber output. It is straightforward to correct for this by replacing the plano-convex MLA with something similar to the biconvex MLA seen in Figure~\ref{fig:mla}. Consequently pupil-transfer systems will have similar, but simpler descriptions and constraints compared to the image-relay microlens systems. For this reason we break out the constraints from both the bi-convex and plano-convex elements.

\subsection{The merit of telecentric designs}
\label{sec:telecentric}

While the re-imaging MLA system does not require additional air-glass surfaces to create a telecentric design, it does place constraints on the lenslet design as we will see below. Further, as noted above a telecentric pupil-imaging MLA system does require the addition of two air-glass surfaces. Broadly, then it is worth considering whether imposing telecentric designs are worth the cost in either design constraints or reflection losses. It is easiest to understand this in the context of the pupil-imaging MLA system as follows.

As a general, qualitative statement for the telecentric pupil-imaging MLA case, given modern multi-layer anti-reflection coatings with broad-band performance achievable at 0.5-1\% reflection loss per air-glass surface over more than an octave in wavelength, the bar is not set very high to gain by adding a two air-glass surfaces to eliminate gFRD.

\begin{figure}[h]
\centering
\includegraphics[width = \linewidth]{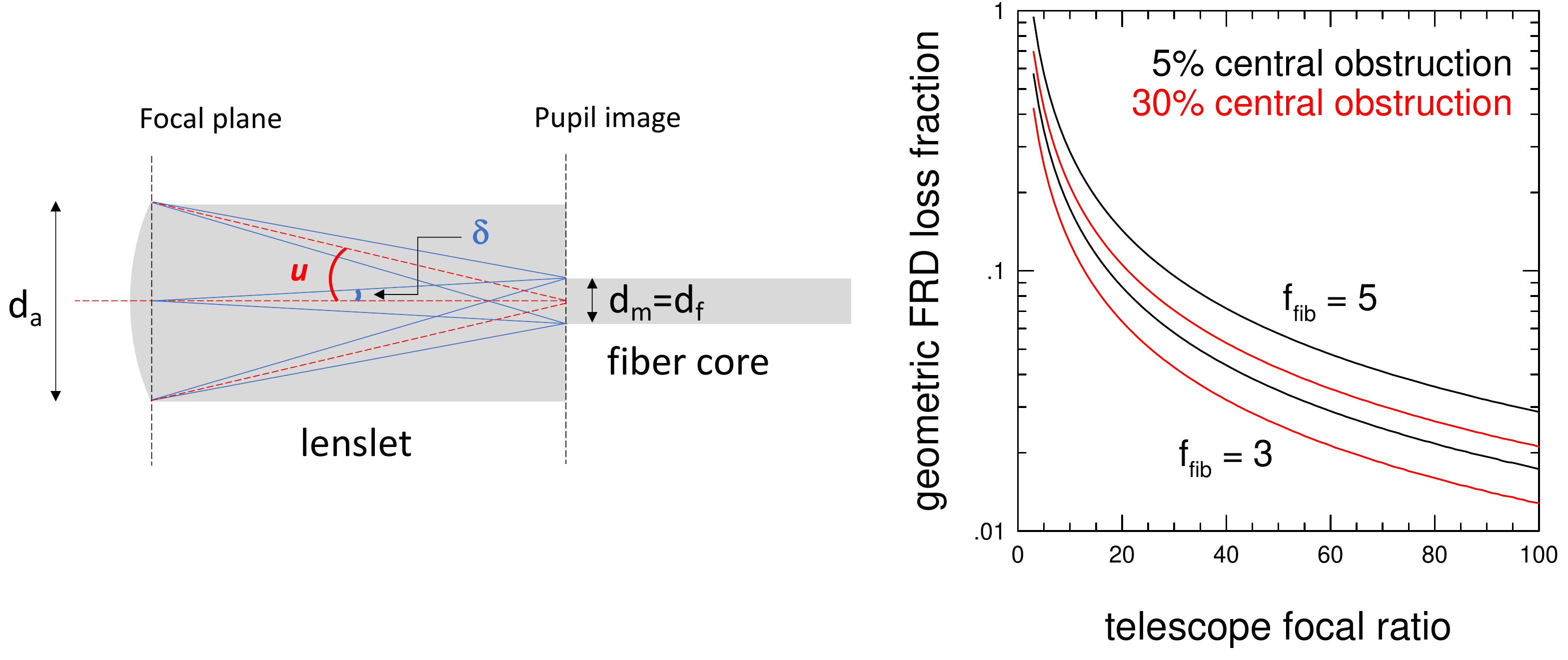}
\caption{Ray schematic for the \textit{non}-telecentric PC MLA pupil-imaging fiber system (left). The telescope focal plane is formed at the location of the left-most vertical dashed line. Blue rays trace the telescope pupil formed by the lenslet and also the fiber-input light cones at the edge of the fiber. The chief ray of these cones is at an angle $\delta$ with respect to the telecentric angle. Red rays (dashed) trace the input cone on the fiber face at the field center, with a marginal ray angle $u$. The geometric FRD loss based on \cite{Wynne_1993} is shown at right for two telescope central obstructions of 5\% (black) and 30\% (red), and
two fiber input beam speeds (f/3 and f/5 in air, bottom and top curves, respectively for each color). The lenslet and fiber index of refraction is assumed to be matched at $\rm n_g=1.46$.}
\label{fig:pcmaltel}
\end{figure}

Equation 3 of \cite{Wynne_1993} can be used to compute the throughput loss due to gFRD. The fraction of the light lost is given by $L\sim (0.4/B)(\delta/u)$ where $u$ is the angle of the marginal ray with respect to the apex of the telecentric beam, $\delta$ is the angle of the marginal ray in the non-telecentric beam with respect to the apex of the telecentric beam at the field edge, and $B$ is the fractional area of the telescope primary mirror unobscured by the secondary mirror. The angles in a PC MLA pupil-imaging system are illustrated in Figure~\ref{fig:pcmaltel}. As two representative extrema, $B$ has a value of about $0.95$ for the Gemini 8m telescope \cite{Gillett_1996} and a value of about 0.7 for the Sloan 2.5m telescope \cite{Gunn_2006}.

Figure~\ref{fig:pcmaltel} shows that gFRD already leads to 10-30\% throughput loss for $f_{\rm tel}=10$ (depending on fiber injection speed and telescope central obstruction), and this increases rapidly for faster telescopes. For any plausible telescope f-ratio, uncorrected gFRD losses never fall below 1\%. This makes it clear that the telecentric design is preferable in all relevant cases for pupil-imaging. For reference the GMOS IFU on Gemini \citep{Allington-SMith_2002} has a PC MLA lenslet system fed at f/50 and injecting light into the fibers at f/4. This system should have $\sim$5\% thoughtput loss due to gFRD.

\subsection{Critical design parameter space for image-relay microlens systems}
\label{sec:critdesign}

The primary design parameters that have bearing on manufacturing limitations include the surface curvature given by $\rm r_p$ and $\rm r_b$, and the MLA thickness $\rm D_b$ or $\rm D_p$. Based on discussion with, e.g. A$\upmu$S\footnote{Advanced Micro-Optic Systems, https://www.amus.de}, we specify these fabrication-based limits for these parameters:

\begin{enumerate}
 \item Maximum manufacturing limitation on the thickness ($\rm D_b$ or $\rm D_p$) of a microlens of 10 mm and a minimum of 0.3/0.6~mm for PC and BC MLA respectively. 
 \item Fabrication limitation on the minimum radius of curvature of a microlens ($\rm r_b$ or $\rm r_p$) of 0.02 mm.
\end{enumerate}

The specific values for these parameters depend on the fiber input f-ratio $\rm f_{\rm fib}$ and micro-image diameter $\rm d_m$ as well as $\rm f_{\rm tel}$. We also bound $\rm f_{\rm fib}$ and $\rm d_m$ in our analysis, as follows, leaving $\rm f_{\rm tel}$ unbounded. We constrain $\rm d_m$ by the range of off-the-shelf multi-mode fiber size, with 50 $\rm \upmu$m as the smallest, and 600 $\rm \upmu$m as the largest that has seen practical use \cite{Eigenbrot18}. 

The range of $\rm f_{\rm fib}$ depends on the range of acceptable f-ratios for the spectrograph input (the collimator).  Traditional spectrograph designs have preferred slow collimators to increase spectral resolution by decreasing the angular size of the slit while keeping the overall beam size as small as possible. Slower input beams offer the added opportunity of demagnification via fast camera optics to optimize matching to detector focal-planes. However, with fiber coupling, FRD requires collimator speeds of f/5 to prevent significant loss of etendue, as seen in successful fiber-fed spectrograph designs (Hectoechelle\cite{hectoechelle}; GIRAFF\cite{giraffe}; upgraded WIYN Bench\cite{mab2008}; SDSS/BOSS\cite{boss}). Indeed, in more recent designs (e.g.,  DESI-f/3.7\cite{desi}, M2FS-f/3.4\cite{m2fs}, WEAVE-f/3.1\cite{weave}, MEGARA-f/3\cite{megara}, PFS-f/2.8\cite{pfs}, DOTIFS-f/4.5\cite{Haeun}), collimators are substantially faster to further minimize FRD. Spectrograph collimators are able to accept f/2.5 with standard off the shelf optics while custom modification may able to grasp faster beams\cite{amase} approaching the fiber numerical aperture (typically f/2.2 for standard, high-transmission broad-bandpass step-index fiber). We take $\rm 2.5<f_{\rm fib}<5$ to be a reasonable range for this parameterin order to keep FRD-induced entropy increase minimal to modest (see Figure~\ref{fig:frd}) in standard step-index fiber optics with numerical apertures of 0.22, most commonly used in astronomical instrumentation. 

In summary, then:

\begin{enumerate}
\setcounter{enumi}{2}
 \item Fiber diameter $\rm d_m$ between 50 to 600 $\rm \upmu$m.
 \item Fiber input beam speed $\rm f_{\rm fib}$ between f/2.5 and f/5.
\end{enumerate}

\section{Analytic model of micro-lens optical parameters}
\label{sec:params}

In what follows it is useful to define the parameter $\rm \upeta$ as the ratio of telescope to fiber focal-ratios:
\begin{equation}
\upeta \ \equiv \ \rm f_{\rm tel} / \rm f_{\rm fib}.
\end{equation}
Appendix~\ref{app:derivation} then presents derivations of
equations for all of the microlens design parameters 
(surface curvatures $\rm r_b,r_p$; thickness and spacing $\rm D_b, D_G, D_p$) as a function of the telescope input and fiber input focal-ratios, the size of the micro-image ($\rm d_m$) and the glass index $\rm n_g$. As before, we equate the micro-image ($\rm d_m$) with the fiber core diameter.

For a fiber-microlens based IFU, the fiber radius can be chosen based on the camera and detector size and usually ranges from 50 to 600 $\rm \upmu$m. Thus for each $\rm d_m$ and $\rm f_{\rm fib}$ we should be able to find a $\rm f_{\rm tel}$ where the microlens radius of curvature is manufacturable while providing adequate clear aperture as well as acceptable spot size even on the edge field. At this point we can vary the telescope f-ratio and the ``aperture ratio" of the beam diameter at the PC array ($\rm d$) to the available aperture diameter ($\rm d_a$). All these quantities are shown in the Figure~\ref{fig:mla}.

% The model is described by the following equations which are derived in detail in Appendix \ref{app:derivation}: 

% \begin{equation}
%    \ \upeta \ \equiv \ \frac{\rm f_{\rm tel}}{\rm f_{\rm fib}},
%\end{equation}

% n\begin{equation}
% \label{eq:da}
%  \rm d_a \ = \rm d_m \ \upeta,
% \end{equation}

% \begin{equation}
% \label{eq:d}
%  \rm d \ =  \rm d_m \  (\upeta - 1),
% \end{equation}

% \begin{equation}
% \label{eq:Dp}
% \rm D_p \ = \rm d_m \rm n_g \rm f_{\rm fib} \ (\upeta - 1),
% \end{equation}

% \begin{equation}
% \rm r_p \ = \rm {d_m} \rm f_{\rm fib} \rm (n_g - 1) (\upeta - 1), 
% \label{eq:rp}
% \end{equation}

% \begin{equation}
%  \rm D_G \ = \rm {d_m} \rm f_{\rm fib} \ (\upeta - 1),
%  \label{eq:Dg}
% \end{equation}

% \begin{equation}
% \rm D_b \ = \rm d_m \rm n_g \rm f_{\rm tel}  \ (\upeta - 1),
% \label{eq:Db}
% \end{equation}

%\begin{equation}
% \rm r_b \ = \rm d_m \ \upeta \left[\left(\sqrt{1/4+(\rm n_g \ \rm f_{\rm fib} \ (\upeta-1))^2} - \rm f_{\rm fib} \ (\upeta-1)\right)^2+1/(4{\rm n_g^2})\right]^{1/2}. 
% \label{eq:rb2}
%\end{equation}

\section{Model Considerations: On-Axis Aberrations}
\label{sec:abblimits}

The dominant on-axis aberrations in the micro-lens optical system are due to sphericity and diffraction. Minimizing these as well as higher-order off-axis aberrations is desirable to maximize the encircled energy injected into a finite fiber core. While our analytical model can deal with geometric aberrations (such as SA), the effort to minimize these depends on whether they dominate. First we show that diffraction, even for micro-lens systems, remains a minimal effect, which allows us to focus on the impact of minimizing SA on the lenslet design. These calculations are done at an effective wavelength of 800~nm which is the logarithmic mid-point for good fiber transmission (400-1600nm), and the worse-case limit for diffraction in visible-wavelength systems.

\begin{figure}[h]
\centering
\includegraphics[width = \linewidth]{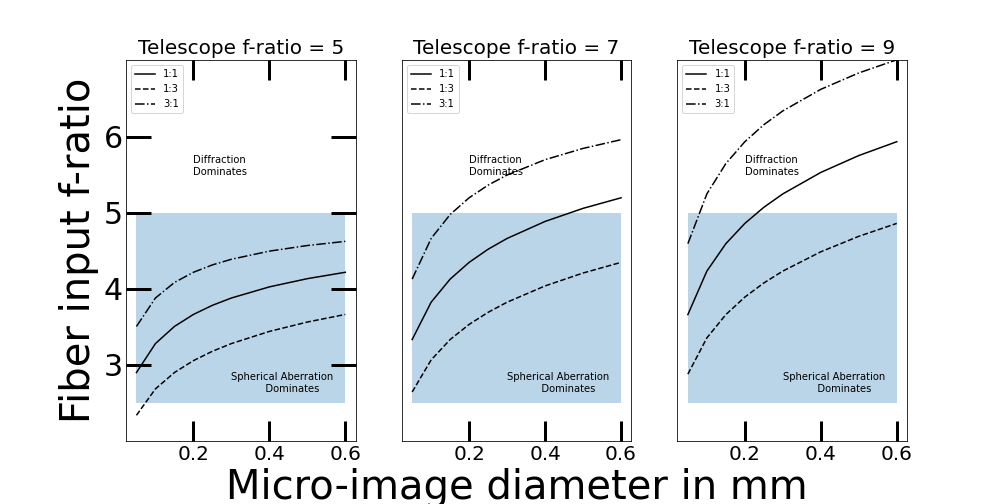}
\caption{Range of fiber input beam speed and micro-image diameter (matched to the fiber core diameter)} for which diffraction or spherical aberration dominates for different telescope beam speeds. For practical purposes, the fiber core diameter will be close to the micro-image diameter. Curves indicate the ratio of the airy disk first minimum for a wavelength of 800~nm to the circle of least confusion at values of unity (solid line), 1:3 (dashed; strongly spherical-aberration limited) and 3:1 (dot-dash; strongly diffraction-limited). The aperture and radius of curvature are computed using equations \ref{eq:d} and \ref{eq:rp} from Appendix~\ref{app:derivation}. The blue shaded region is the design target for fiber input f-ratio.
\label{fig:diffVtsa}
\end{figure}

To compare diffraction to SA we compare the impact on  the radius enclosing 90\% of the encircled energy (EE90). In case of SA, the effect can be estimated by the transverse component i.e. the radius of circle of least confusion (CoC) which can be roughly estimated (in radians) as $\rm d^3/( 16\rm r^3)$, where d is the diameter of the entry aperture and r is the radius of curvature. Using equations \ref{eq:d} and \ref{eq:rp} for $\rm d$ and $\rm r_p$ from Appendix~\ref{app:derivation}, we can express this as $1/(16 {\rm f_{fib}}(\rm n_g - 1))^3$. The increase in edge field EE90 due to diffraction is simply the first minimum of the airy disk which is measured as $1.22 \uplambda/\rm d$ radians. For a system which converts an f/11 telescope beam into an f/3 fiber input within a microimage diameter of 100$\upmu$m, the values of $\rm r_p$ and d are 0.365 mm and 0.267 mm respectively. At 800 nm, the EE90 due to SA is more than 6 times greater than the same for diffraction. Figure \ref{fig:diffVtsa} shows the range of $\rm f_{fib}$ and $\rm d_m$ (note again: we are assuming $\rm d_m$ is the same as the fiber core diameter) where diffraction or SA dominates for different telescope beam speed. It is interesting to note that the system tend to be more diffraction limited with slower $\rm f_{fib}$ for a given $\rm d_m$ and $\rm f_{tel}$. Slower $\rm f_{fib}$ would also (1) introduce additional FRD; (2) reduce the grasp of the system, (3) reduce the radius of curvature of the optical elements; and hence are not advisable to use. Thus for practical purpose we would concentrate on the range of $\rm f_{fib}$, $\rm d_m$ and $\rm f_{tel}$ where SA dominates.\textit{ Our f-ratio selection range also shows that beyond $\rm f_{tel}/7$, most of our optimization space comes under systems that are mostly SA dominated.}

\section{Constraining the telescope focal-ratio: minimizing throughput loss due to aberrations}
\label{sec:minloss}

Given the dominance of SA over diffraction in most applications, we use our analytic model to minimize the effect of spherical aberration on diminishing throughput. We then check if this analytic model is accurate when considering higher-order aberrations, and conclude on the limits our calculations place on the telescope focal-ratio. 

To minimize the effects of light-loss from SA, we assume the thickness of microlens positions the CoC on the exit surface of the PC microlens (the fiber input core location). We can estimate this PC thickness as the distance from the entry into MLA to the point of crossing the optical axis by a marginal ray given the separation between the chief and the marginal ray and lenslet radius of curvature is known. This is an analytical description of the longitudinal SA. We have used this description to derive the relative constraint between PC MLA thickness and radius of curvature for a given $\rm d_a$ and $\rm d_m$ (hence $\rm \upeta$). Likewise for a given radius of CoC (depending on how fast is the PC MLA) we can work out the losses for a fiber with a diameter of $\rm d_m$. If the prescription on the PC MLA thickness and radius become too severe we can define the trade-off in photon loss outside of $\rm d_m$ if the CoC is {\it not} placed on the exit surface of the PC MLA.

Appendix \ref{ap:focal length} and \ref{ap:TSALSA} describe the analytical model to define the longitudinal and transverse SA (LSA and TSA). We found that analytical solutions to be cumbersome to define the location and radius of CoC and hence we used the numerical approach. It may be noted that the result of the analytical model is still important as an initial condition for performing optical simulation to predict precise SA losses in specific IFU design studies.

We used d/2 as the marginal ray height and normalized with $\rm r_p$. Their ratio only depends on $\rm f_{fib}$ and $\rm n_g$. On the other hand, the normalized paraxial focus location is $\rm n_g/(n_g-1)$ as described in Appendix~\ref{ap:focal length}. Given this, we compute the radius and location of CoC and marginal focus for a grid of $\rm r_2/r_p$ (refer to Figure~\ref{fig:sa} in Appendix~\ref{ap:TSALSA}) from 0 to 0.15 with 101 samples while the position along the optical axis had 51 samples between 3 and 3.3~mm. The result is tabulated in Table \ref{tab:spab} which establishes the well known facts (\citenum{schroeder}): (a) the radius of the circle of least confusion, $\rm r_{coc}$, is $\sim$ 0.25 times the beam radius at marginal ray height; (b) the distance of CoC from paraxial focus is $\sim$ 0.75 times the distance of marginal focus from the paraxial focus. The minor variation in TSA and LSA due to fiber input beam speed is negligible.

\begin{table}[h]
\scriptsize
\centering
\begin{tabular}{@{}cccccccc@{}}
\toprule
$\mathbf{f_{fib}}$ &
  \textbf{pf} &
  \textbf{mf} &
  \textbf{coc} &
  $\mathbf{\frac{pf-coc}{pf-mf}}$ &
  $\mathbf{r_{coc}}$ &
  $\mathbf{r_{pf}}$ &
  $\mathbf{\frac{r_{coc}}{r_{pf}}}$ \\ \midrule
2.5 & 3.19 & 3.04  & 3.075  & 0.76 & 0.0056 & 0.0223 & 0.25 \\
3   & 3.19 & 3.088 & 3.1125 & 0.76 & 0.003  & 0.0124 & 0.24 \\
3.5 & 3.19 & 3.116 & 3.135  & 0.74 & 0.0019 & 0.0076 & 0.25 \\
4   & 3.19 & 3.133 & 3.1475 & 0.75 & 0.0013 & 0.005  & 0.26 \\
4.5 & 3.19 & 3.146 & 3.157  & 0.75 & 0.0009 & 0.0034 & 0.26 \\
5   & 3.19 & 3.154 & 3.1625 & 0.76 & 0.0007 & 0.0025 & 0.28 \\ \bottomrule
\end{tabular}
\caption{Effect of fiber input beam speed on the location and radius of circle of least confusion. From left to right, 
$\rm f_{fib}$ is the fiber injection speed, pf is the paraxial focal length, mf is the focal length from the marginal ray, coc is the distance to the circle of least confusion, $\rm r_{coc}$ is the radius of the circle of least confusion, and $\rm r_{pf}$ is the beam radius at the paraxial focus. All the quantities are normalized by the plano-convex radius of curvature. All dimensions are in mm except for ratios and f-number.}
\label{tab:spab}
\end{table}

\subsection{Accuracy of the analytic model}
\label{sec:accuracy}

At this point it is important to note that there is also SA as well as off-axis aberration introduced by the biconvex lens entry and exit surfaces. The analytical optimization of multi surface SA is non-trivial and hence Zemax is used for this purpose. We used optical parameters from our analytic model as initial conditions for the Zemax optimization and looked for how these parameters change after detailed optimization. This optimization accounts for higher order aberration such as coma, astigmatism, color etc. We used two configurations, one each in regimes where SA or diffraction dominate (refer to Figure~\ref{fig:diffVtsa}): 

\begin{enumerate}
 \item SA dominated regime: a f/11 telescope beam is relayed to a f/3.7 fiber input beam for a micro-image diameter of 100$\upmu$m.
 \item Diffraction-limited regime: a f/7 telescope beam is relayed to a f/4 beam for a micro-image diameter of 50 $\upmu$m.
\end{enumerate}

We use equations \ref{eq:da} to \ref{eq:rb2} for $\rm d_a$ and $\rm r_b$ from Appendix~\ref{app:derivation} to predict the optical parameters of the above mentioned configurations. The analytical model is built on the assumption that the pupil should be formed at the exit surface of the BC MLA and the PC MLA is going to form the CoC at its exit surface. We then use Zemax optimization to further optimize both the configurations. Table \ref{tab:comp} shows the comparison chart of the parameters and performance before and after optimization. The design parameters change little in the SA-limited case (as we might expect), but significantly in the diffraction-limited regime. Nonetheless, we find that the analytical prediction produces optical \textit{performance} similar to that of Zemax optimized design within 4\% for both the SA dominated regime \textit{and} the diffraction dominated scenario based on EE98 estimates. Since the fiber azimuthally scrambles the spatial information at its input, RMS spot radius do not provide the required estimate of performance.

\begin{table}[]
\scriptsize
\centering
\begin{tabular}{@{}llcccc@{}}
\toprule
 &
  \textbf{\begin{tabular}[c]{@{}c@{}}dimensions\\ are in $\upmu$m\end{tabular}} &
  \multicolumn{2}{c}{\textbf{\begin{tabular}[c]{@{}c@{}}Spherical Aberration\\ dominated\end{tabular}}} &
  \multicolumn{2}{c}{\textbf{\begin{tabular}[c]{@{}c@{}}Diffraction\\ dominated\end{tabular}}} \\ \midrule
\multirow{3}{*}{\textbf{Configuration}} &
  (a) $f_{tel}$ &
  \multicolumn{2}{c}{11} &
  \multicolumn{2}{c}{7} \\
                                            & (b) $f_{fib}$ & \multicolumn{2}{c}{3.7} & \multicolumn{2}{c}{4}  \\
                                            & (c) $d_m$     & \multicolumn{2}{c}{100} & \multicolumn{2}{c}{50} \\ \midrule
\textbf{} &
   &
  \textbf{Model} &
  \textbf{Simulation} &
  \textbf{Model} &
  \textbf{Simulation} \\ \midrule
\multirow{5}{*}{\textbf{Optical parameter}} & (d) $r_p$     & 333        & 327        & 68.4        & 114      \\
                                            & (e) $D_p$     & 1050       & 994        & 218.4       & 305      \\
                                            & (f) $D_G$     & 730        & 469        & 150         & 103      \\
                                            & (g) $r_b$     & 999        & 887        & 126         & 141      \\
                                            & (h) $D_b$     & 3160       & 3119       & 382         & 578      \\ \midrule
\multirow{3}{*}{\textbf{Performance}}       & (i) $d_{f0}$  & 1.207      & 1.121      & 0.317       & 0.4      \\ 
                                            & (j) $d_{f1}$  & 3.424      & 2.919      & 3.555       & 1.24     \\
                                            & (k) EE98      & 52         & 50         & 24.5        & 25       \\ \bottomrule
\end{tabular}
\caption{Comparison of design parameters and performance of the reimaging microlens system based on our analytic model (Model) and after optimization using Zemax (Simulation) for different configurations in spherical aberration (SA) dominated and diffraction dominated case. The different configuration, optical and performance parameters are: (a) $\rm f_{tel}$: telescope focal ratio; (b) $\rm f_{fib}$: fiber input focal ratio; (c) $\rm d_m$: micro-image diameter at the fiber input face; (d) $\rm r_p$: plano-convex microlens radius of curvature; (e) $\rm D_p$: plano-convex microlens thickness; (f) $\rm D_G$: air gap thickness between plano-convex and bi-convex microlens; (g) $\rm r_b$: bi-convex microlens radius of curvature; (h) $\rm D_b$: bi-convex microlens thickness; (i) $\rm d_{f0}$: RMS spot size at the field center of the fiber input surface; (j) $\rm d_{f1}$: RMS spot size at the field edge of the fiber input surface; (k) EE98: radius of 98\% of the normalized encircled energy at the fiber input surface. This corresponds to the reimaged patch of $d_a$ = $d_m \times \upeta$. All dimensions, except for focal ratios, are provided in $\upmu$m. Model and Simulation refer to analytical model predicted value and Zemax optimization provided value respectively.}
\label{tab:comp}
\end{table}

\subsection{Limiting telescope beam speed in the presence of micro-lens aberrations}
\label{sec: limits}

For a single spherical lens system (e.g., just the PC MLA), intuitively one would place the CoC produced by the lens on the fiber input aperture, which is what our analytic model assumes. For a compound lens system the situation is less obvious, however our Zemax optimization indicates (Table \ref{tab:comp}) that placing the CoC produced by the PC MLA very close to the fiber entrance aperture remains optimal when spherical aberration dominates over diffraction on-axis.

\begin{figure}[H]
\centering
\includegraphics[width = 0.8\linewidth]{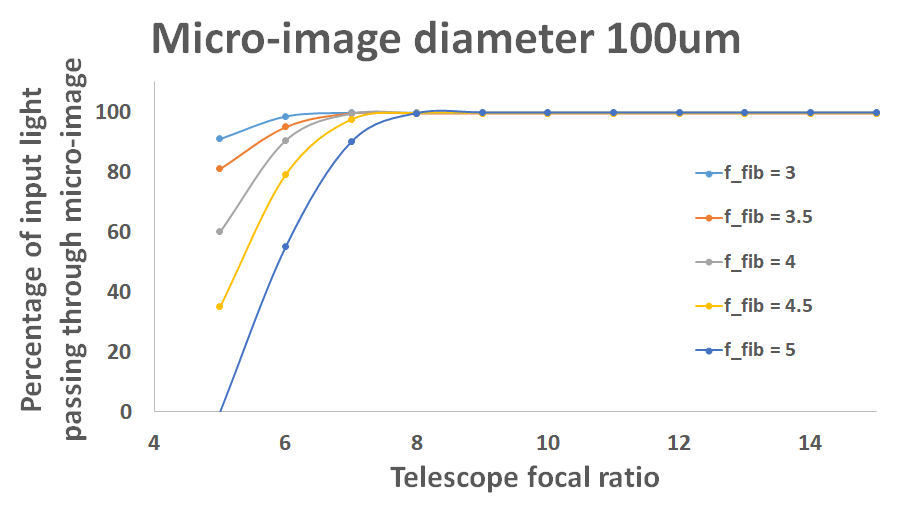}
\caption{Percentage of throughput captured within the micro-image diameter ($\rm d_m$, taken to be the same as the fiber diameter) against telescope focal-ratio for different fiber input beam speed for plano-convex MLA assuming collimated input beam in the presence of diffraction, spherical aberration and all higher-order aberrations. Results are based on Zemax modeling. Similar results are found for micro-image diameters ranging from 50 to 600 $\rm \upmu$m.}
\label{fig:tp100}
\end{figure}

We have used Zemax simulation to define the photon loss for a fiber capturing the CoC at its entry surface for a range of telescope and fiber input f-ratio. The throughput (provided as percentage of input photons) captured within the micro-image diameter of 100 $\rm \upmu$m for different $\rm f_{\rm tel}$ and $\rm f_{\rm fib}$ is plotted in figure \ref{fig:tp100} including the effects of diffraction and off-axis aberrations. The Figure shows that $\rm f_{\rm tel}$ of 6, 7, 7, 8 and 8 are optimum for $\rm f_{\rm fib}$ of 3, 3.5, 4, 4.5 and 5 respectively. Although the plot is for 100 $\rm \upmu$m microimage diameter, the optimum telescope f-ratio of f/8 or slower holds true for other micro-image diameters ranging from 50 to 600 $\rm \upmu$m. This discussion can be used as a constraint on $f_{tel}$ coming from aberrations and can be summarized as the following:
\begin{enumerate}
    \item The faster the telescope beam, the larger the SA, {\uline as well as off-axis aberrations}, hence the larger the losses.
    \item At fast telescope beams, a fast fibre input beam helps to minimize the losses.
\end{enumerate}

\section{Constraining the telescope focal-ratio: microlens manufacturing limitations} 
\label{sec:constraints}

Aberrations are not the only component that set a bound on the usable telescope focal ratio, $\rm f_{\rm tel}$. Several optical parameters, such as the radius of curvature and thickness of PC and BC microlenses also pose constraints. In this section we describe the limiting conditions originating from such manufacturing limitations for various micro-image diameters, $\rm{d_m}$, and fiber injection speeds, $\rm {f_{fib}}$. These limits are computed using our analytical model in Section~\ref{sec:params}. We use openly available information provided by A$\upmu$S as a benchmark for fabrication.

\subsection{Limits on \texorpdfstring{$f_{tel}$}{TEXT} from lenslet radii of curvature}
\label{sec:roc}

\begin{figure}[H]
\centering
\includegraphics[width = 0.71\linewidth]{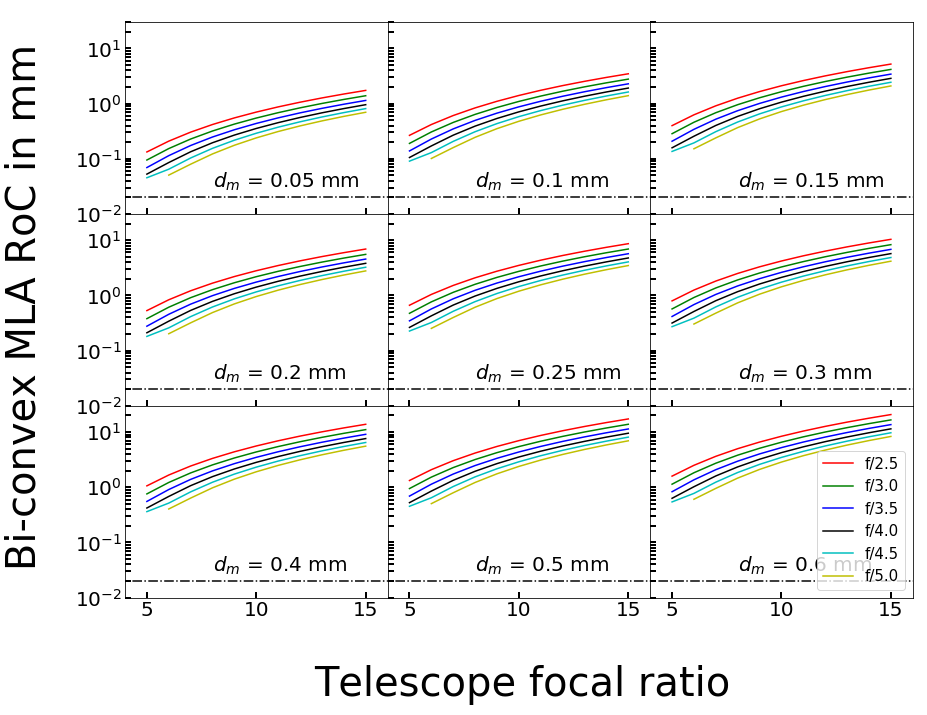}
\caption{Dependence of bi-convex microlens (BC) radius of curvature ($\rm r_b$) on telescope f-ratio for a range of fiber input f-ratios (different colored lines defined the bottom-right panel key) and fiber diameters, assumed here to equal to $\rm d_m$. From top-left to bottom right panels correspond to $\rm 50~\upmu$m to $\rm 600~\upmu$m core fiber. Black dashed lines denote 0.02 mm, the \textit{minimum} possible radius of curvature.}
\label{fig:rb}
\end{figure}

\begin{figure}[H]
\centering
\includegraphics[width = 0.71\linewidth]{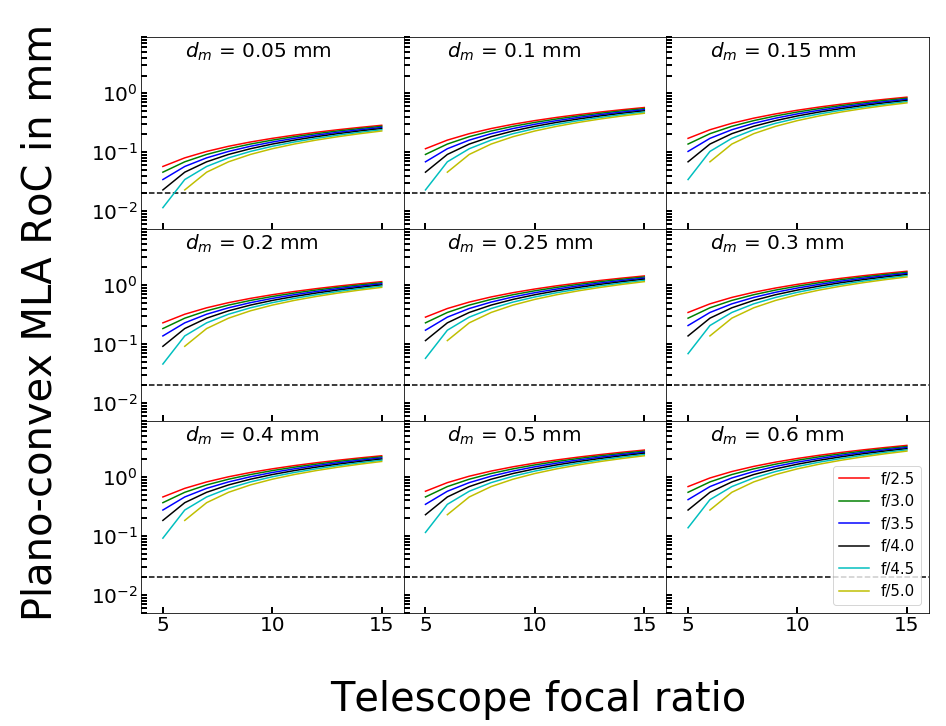}
\caption{Dependence of plano-convex microlens (PC) radius of curvature ($\rm r_p$) on telescope f-ratio, fiber diameter, and fiber input f-ratio as in Figure~\ref{fig:rb}. Curves are defined as in Figure~\ref{fig:rb}.}
\label{fig:rp}
\end{figure}

Figures \ref{fig:rb} and \ref{fig:rp} show the effect of $\rm f_{\rm fib}$, and $\rm d_m$ and $\rm f_{\rm tel}$ on the required radii of curvature (RoC) for bi-convex ($\rm r_b$) and plano-convex ($\rm r_p$) MLAs, respectively. Clearly the fabrication limit on $\rm r_b$ does not constrain $\rm f_{tel}$ for any fiber size and injection speed consider.  Similarly, for On the other hand, the fabrication limit on $\rm r_p$ shows that for $\rm 50~\upmu$m fiber it is necessary for $\rm f_{\rm tel} \geq 6$ for $\rm f_{\rm fib}=5$, but it does not provide any restriction for faster fiber injection speeds $\rm f_{\rm fib}<4$.  For larger fibers ($\geq100\upmu$m) we can conclude broadly that there is no limiting $\rm f_{\rm tel}$ originating from fabrication constraints on the MLA RoC.

\subsection{Limits on \texorpdfstring{$f_{tel}$}{TEXT} from MLA thickness}
\label{sec:thickness}

\begin{figure}[h]
\centering
\includegraphics[width = 0.71\linewidth]{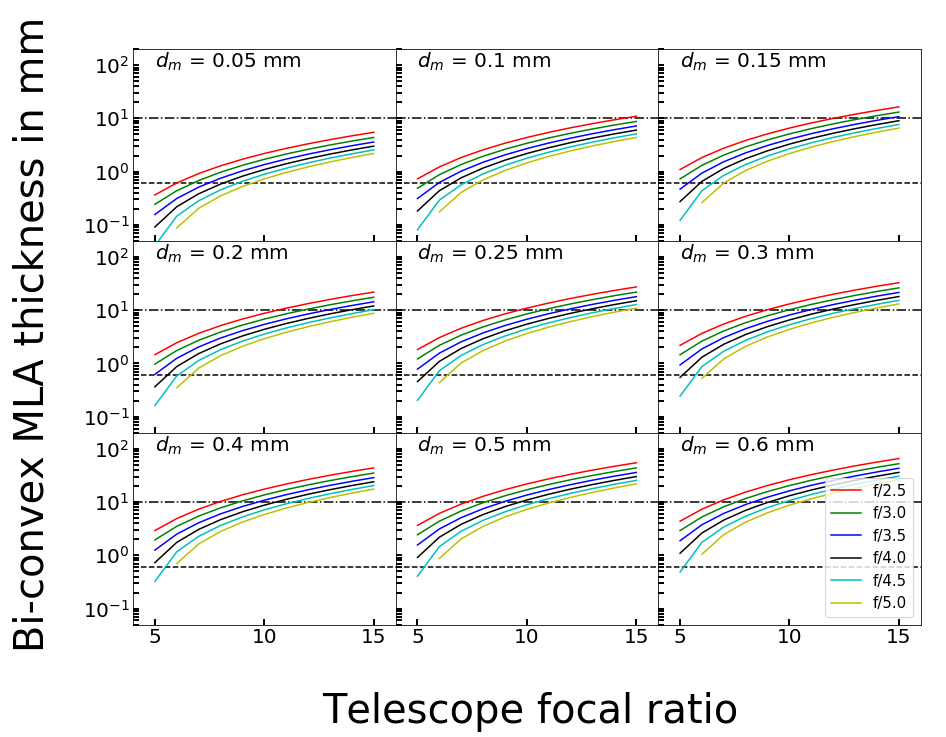}
\caption{Dependence of bi-convex microlens required thickness ($\rm D_b$) on telescope f-ratio, fiber diameter $\rm d_m$, and fiber input f-ratio (different colored lines defined the bottom-right panel key). The dash-dot and the dashed black lines represent the higher and the lower limit of thickness respectively.}  
% It should be noted that with additional effort of alignment and gluing, two PC MLAs and arbitrarily thick glass plate can be used to eliminate the thickness upper limit.}
\label{fig:db}
\end{figure}

\begin{figure}[h]
\centering
\includegraphics[width = 0.71\linewidth]{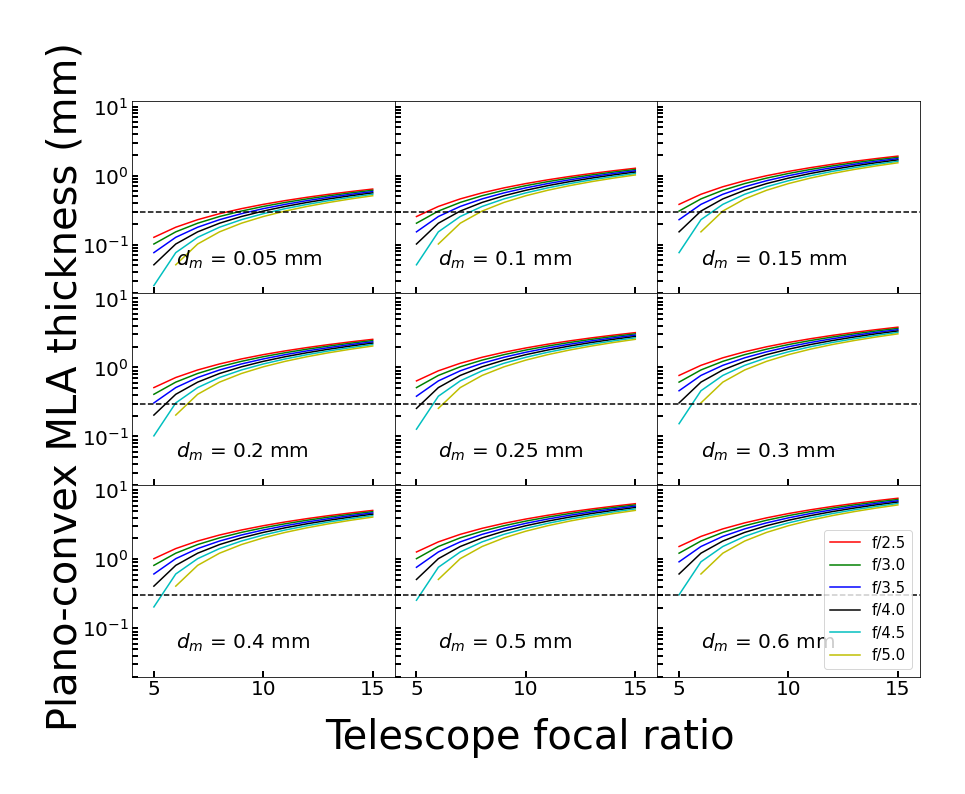}
\caption{Dependence of plano-convex microlens  required thickness $\rm D_p$
on telescope f-ratio, fiber diameter $\rm d_m$, and fiber input f-ratio (different colored lines defined the bottom-right panel key). The dashed black lines represent the thickness lower limit; the upper limit is at the upper limit of the plotting range.}
\label{fig:dp}
\end{figure}

Apart from the microlens radius of curvature, the microlens thicknesses ($\rm d_p$ and $\rm d_b$) also constrain the telescope f-ratio range. Equations \ref{eq:dp} and \ref{eq:db} in Appendix~\ref{app:derivation} show both $\rm d_p$ and $\rm d_b$ increase linearly with $\rm d_m$ and hence the manufacturable thickness could pose constraints on $\rm f_{\rm tel}$. We have computed the dependence of $\rm f_{\rm tel}$ on MLA thickness for different fiber injection speeds and micro-image diameters in Figures \ref{fig:db} and \ref{fig:dp}.

For fabrication thickness \textit{upper} limits, a PC MLA can have arbitrarily large thickness by bonding it to a flat substrate. However, for the BC MLA, the registration error between curved faces increases with thickness and hence a manufacturing limit of 10~mm is used by the vendors. It should be noted that with additional effort of alignment and bonding, two PC MLAs and arbitrarily thick glass plate can be used to eliminate the thickness upper limit. Consequently the BC MLA thickness upper limit is taken to be a soft limit. In general MLA thickness upper limits tend to drive the telescope speed to be faster. As can be seen in Figures \ref{fig:db} and \ref{fig:dp}, a 10~mm thickness upper limit, the bi-convex MLA sets $\rm f_{\rm tel} \leq 7 (11)$ for $\rm f_{\rm fib} = 2.5 (5)$, respectively, for $\rm 600~\upmu$m fibers. For smaller fibers with diameter $\leq \rm 100~\upmu$m the thickness does not limit the $\rm f_{tel}$ within the range of telescope focal ratios that is being considered. The PC MLA thickness imposes no limits on $\rm f_{\rm tel}$ as it can be increased arbitrarily.

For fabrication thickness \textit{lower} limits, the manufacturer has to ensure that the array is flat and thus it is difficult to reduce lenslet thickness below 0.3 and 0.6~mm for PC and BC MLA respectively. In general MLA thickness lower limits tend to drive the telescope speed to be slower. As can be seen in Figures \ref{fig:db} and \ref{fig:dp}, the minimum thickness forces $\rm f_{tel} \geq 9 (11)$ for 50~$\upmu$m fiber for $\rm f_{fib} = 2.5 (5)$ but does not limit fibers with $\geq 400 \upmu$m cores.

\subsection{Summary of limitations on \texorpdfstring{$\rm f_{tel}$}{TEXT}}
\label{sec:summlim}

Table \ref{tab:summary} compiles all the limitations on $\rm f_{\rm tel}$ from different criteria presented in this and the previous section (\ref{sec:abblimits}). In the final columns we take the most conservative option as our final limit. In most cases the lower limit is set by `SA,' which includes the impact of all aberrations on EE98. This means that this limit is unaffected by any potential future improvements in fabrication limits for spherical lenslet figures. Aspherics may alter this picture, and are worthy of future consideration. In constrast, the upper limit on $\rm f_{tel}$ is constrained solely by the BC MLA thickness upper limit. As described above, this upper limit may be surmountable and as such there is no hard upper limit on the telescope beam speed.

It is evident and intuitive that faster telescope beams require the optical dimensions of the microlens arrays to be smaller and hence lossier for fiber-coupling due to increased aberrations. Less intuitive is the fact that for larger fiber core diameter, the limiting telescope f-ratio gets faster. The effect is most pronounced going from  the diffraction-limited to SA dominated regimes (in the range $\rm 50~\upmu m < d_m < 100~\upmu m$). Indeed, in the SA dominated regime the limiting telescope f-ratio is constant for a given fiber input beam speed, and only changes by 33\% for a factor of two change in fiber injection speed.

\begin{table}[h]
\scriptsize
\centering
\label{tab:my-table}
\begin{tabular}{@{}cccccccccccccc@{}}
\toprule
\multirow{3}{*}{\textbf{$\rm d_m$ ($\upmu$m)}} &
  \multirow{3}{*}{\textbf{$\rm f_{fib}$}} &
  \multicolumn{4}{c}{\textbf{Thickness}} &
  \multicolumn{4}{c}{\textbf{RoC}} &
  \multicolumn{2}{c}{\multirow{2}{*}{\textbf{Abr}}} &
  \multicolumn{2}{c}{\multirow{2}{*}{\textbf{Final}}} \\ \cmidrule(lr){3-10}
 &
   &
  \multicolumn{2}{c}{\textbf{BC}} &
  \multicolumn{2}{c}{\textbf{PC}} &
  \multicolumn{2}{c}{\textbf{BC}} &
  \multicolumn{2}{c}{\textbf{PC}} &
  \multicolumn{2}{c}{} &
  \multicolumn{2}{c}{} \\ \cmidrule(l){3-14} 
 &
   &
  \textbf{L} &
  \textbf{U} &
  \textbf{L} &
  \textbf{U} &
  \textbf{L} &
  \textbf{U} &
  \textbf{L} &
  \textbf{U} &
  \textbf{L} &
  \textbf{U} &
  \textbf{L} &
  \textbf{U} \\ \midrule
\multirow{6}{*}{50}  & 2.5 & 6  & -  & 9  & - & - & - & - & - & 6 & - & 9  & -  \\
                     & 3   & 7  & -  & 9  & - & - & - & - & - & 6 & - & 9 & -  \\
                     & 3.5 & 8  & -  & 10 & - & - & - & - & - & 7 & - & 10 & -  \\
                     & 4   & 9  & -  & 10 & - & - & - & - & - & 7 & - & 10 & -  \\
                     & 4.5 & 9  & -  & 11 & - & - & - & 6 & - & 8 & - & 11 & -  \\
                     & 5   & 10 & -  & 11 & - & - & - & 6 & - & 8 & - & 11 & -  \\ \midrule
\multirow{6}{*}{100} & 2.5 & -  & 14 & 6  & - & - & - & - & - & 6 & - & 6  & 14 \\
                     & 3   & 6  & -  & 7  & - & - & - & - & - & 6 & - & 7  & -  \\
                     & 3.5 & 6  & -  & 7  & - & - & - & - & - & 7 & - & 7  & -  \\
                     & 4   & 7  & -  & 7  & - & - & - & - & - & 7 & - & 7  & -  \\
                     & 4.5 & 8  & -  & 8  & - & - & - & - & - & 8 & - & 8  & -  \\
                     & 5   & 8  & -  & 8  & - & - & - & - & - & 8 & - & 8  & -  \\ \midrule
\multirow{6}{*}{200} & 2.5 & -  & 10 & -  & - & - & - & - & - & 6 & - & 6  & 10 \\  
                     & 3   & -  & 11 & -  & - & - & - & - & - & 6 & - & 6  & 11 \\
                     & 3.5 & -  & 12 & -  & - & - & - & - & - & 7 & - & 7  & 12 \\
                     & 4   & 6  & 13 & 6  & - & - & - & - & - & 7 & - & 7  & 13 \\
                     & 4.5 & 7  & 14 & 6  & - & - & - & - & - & 8 & - & 8  & 14 \\
                     & 5   & 7  & -  & 7  & - & - & - & - & - & 8 & - & 8  & -  \\ \midrule
\multirow{6}{*}{300} & 2.5 & -  & 8  & -  & - & - & - & - & - & 6 & - & 6  & 8  \\  
                     & 3   & -  & 9  & -  & - & - & - & - & - & 6 & - & 6  & 9  \\
                     & 3.5 & -  & 10 & -  & - & - & - & - & - & 7 & - & 7  & 10 \\
                     & 4   & 6  & 11 & -  & - & - & - & - & - & 7 & - & 7  & 11 \\
                     & 4.5 & 6  & 12 & 6  & - & - & - & - & - & 8 & - & 8  & 12 \\
                     & 5   & 7  & 13 & 6  & - & - & - & - & - & 8 & - & 8  & 13 \\ \midrule
\multirow{6}{*}{400} & 2.5 & -  & 7  & -  & - & - & - & - & - & 6 & - & 6  & 7  \\
                     & 3   & -  & 8  & -  & - & - & - & - & - & 6 & - & 6  & 8  \\
                     & 3.5 & -  & 9  & -  & - & - & - & - & - & 7 & - & 7  & 9  \\
                     & 4   & -  & 10 & -  & - & - & - & - & - & 7 & - & 7  & 10 \\
                     & 4.5 & 6  & 11 & 6  & - & - & - & - & - & 8 & - & 8  & 11 \\
                     & 5   & 6  & 12 & 6  & - & - & - & - & - & 8 & - & 8  & 12 \\ \midrule
\multirow{6}{*}{500} & 2.5 & -  & 7  & -  & - & - & - & - & - & 6 & - & 6  & 7  \\ 
                     & 3   & -  & 8  & -  & - & - & - & - & - & 6 & - & 6  & 8  \\
                     & 3.5 & -  & 8  & -  & - & - & - & - & - & 7 & - & 7  & 8  \\
                     & 4   & -  & 9  & -  & - & - & - & - & - & 7 & - & 7  & 9  \\
                     & 4.5 & 6  & 10 & 6  & - & - & - & - & - & 8 & - & 8  & 10 \\
                     & 5   & 6  & 11 & 6  & - & - & - & - & - & 8 & - & 8  & 11 \\ \bottomrule
\end{tabular}
\caption{Summary of limits on telescope beam speed for all configurations of varying micro-image diameter at the fiber input ($\rm d_m$) and fiber input beam speed ($\rm f_{fib}$). L and U denotes lower and upper limits of telescope beam speed respectively while no limits are represented with dash. BC and PC represents bi-convex and plano-convex microlens while Abr and RoC stands for optical aberrations and radius of curvature, respectively. All constraints are from Section~\ref{sec:constraints} except for optical aberrations (Section~\ref{sec:abblimits})}
\label{tab:summary}
\end{table}

\begin{figure}[h]
\centering
\includegraphics[width = 0.9\linewidth]{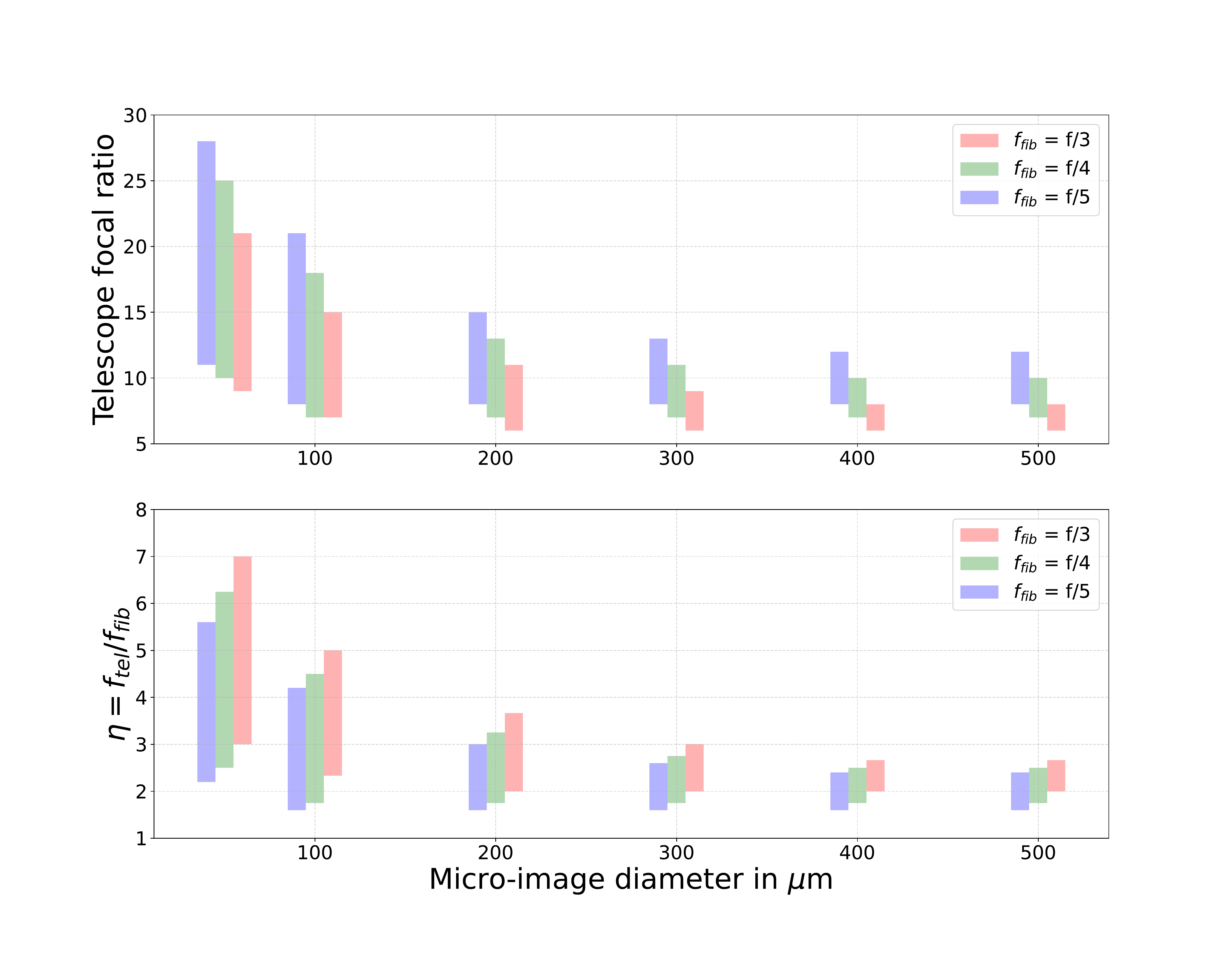}
\caption{Summary of table \ref{tab:summary} in a visual form. Range of acceptable $\rm f_{tel}$ is plotted against different fiber sizes for $\rm f_{fib}$ of f/3, f/4 and f/5.}
\label{fig:tabsum}
\end{figure}

We provide a visual summary of the Table \ref{tab:summary} in Figure~\ref{fig:tabsum} where the final range of acceptable telescope focal ratios are plotted against fiber diameter for three fiber input beam speeds. In terms of conserving grasp, it is important to note that due to FRD it is preferable to use fiber injection speeds closer to  $\rm f_{fib}=3$ than $\rm f_{fib}=5$. Overall, however, it is important to keep in mind that the grasp (A $\Omega$) does not depend on the telescope focal-ratio but rather only on the square of the telescope diameter $\rm D_{tel}$ (collecting area, A), and the product of $\rm{d_m/(D_{tel} \times f_{fib})}$ (solid angle, $\Omega$). While the solid angle can be re-written as $\rm{d_a/(D_{tel} \times f_{tel})}$, fundamentally the grasp is driven by the telescope size and the considerations of the spectrograph injection speed and aperture, hence the fiber size $\rm{d_m}$ and the fiber injection speed.

As a common practice in slit spectroscopy, astronomers often need to modulate the slit width at fixed focal ratio to optimize the resolution-throughput product of their observations given the requirements of the scientific program, the observing conditions, and the angular scale of their sources. For fiber spectroscopy, including lenslet-coupled IFUs, this is equivalent to changing the fiber core diameter at fixed focal ratio. In this context Figure~\ref{fig:tabsum} shows that for fiber sizes smaller than $100~\upmu$m there is a limited range in fiber size that can be accommodated at fixed telescope and fiber injection speeds. Conversely, for fiber core diameters of 100~$\upmu$m and larger lenslet coupling can be achieved with a fixed telescope focal ratio of f/8 for fiber injection at f/3 (or f/9 to f/10 for fiber injection at f/4, etc.).  Similarly, for fiber cores $>100~\upmu$m, there is a broader range of fiber injection speeds at a fixed telescope focal ratio in the range of 8 to 10, with f/8 providing the largest range of fiber injection speeds. On this basis we suggest f/8 as the optimum telescope f-ratio in terms lenslet-fiber coupling flexibility for fibers with core sizes $\geq 100~\upmu$m. The use of smaller fibers may apply to small telescopes (below 4m) where the science requires matching of the fiber core to the seeing disk.  While Figure~\ref{fig:tabsum} indicates that telescope focal ratios faster than f/8 would allow for faster injection speeds for larger fibers, the faster telescope focal ratios will limit the dynamic range in fiber core size over which this is possible.

\section{Summary}
\label{sec:summ}

In this paper we presented the trade-offs in optical design of micro-lens reimaging systems constrained to link telescope focal planes to fiber optics. Specifically, we have examined the constraints on telescope focal ratio for a realistic range of fiber core diameters and fiber injection speeds given (i) manufacturing limitations on micro-lens arrays and (ii) the need to have high-efficiency coupling. For this trade study we have developed an analytical model from first principals and two basic conditions: (a) the BC MLA should produce the pupil at its exit surface; (b) the image produced by the PC MLA must be telecentric to eliminate geometric FRD. We evaluate the efficacy of the model by considering the salient performance parameter of fibers acting as photon buckets: the percentage of energy enclosed with the fiber core by the compound lenslet system. We adopt this metric rather than RMS spot radius since fibers scramble the spatial information. In this regard the performance of the model-predicted optical design matches within 4\% of optimized models using ray-trace software. We have also characterized in detail the different regimes in which these MLA systems are dominated (on axis) by diffraction versus SA. While the detailed micro-lens parameters (thickness, spacing, radius of curvature) are more accurately prescribed by our analytic model in the SA limited regime, in all cases, the model designs provide better than 98\% throughput apart from Fresnel losses at the glass-air boundaries in all regimes.

We computed the overall upper and lower limits on telescope focal ratio set by the fabrication limit of BC and PC MLA as well as light-losses from SA and off-axis aberrations. Lower limits on $\rm f_{tel}$ are set by considering light-losses into the fiber, while upper limits are set by the BC MLA thickness. This upper limit of $\rm f_{tel}$ may be relaxed via additional effort in replacing a monolithic BC MLA with two back-facing PC MLAs, possibly separated by a flat glass plate. 

We find the telescope focal ratio lower-limit falls precipitously from the smallest fibers considered ($50~\upmu$m) to $100~\upmu$m, which corresponds to the transition between diffraction to SA dominated regimes. For fibers larger than $100~\upmu$m the limiting telescope focal-ratio is given roughly by $\rm f_{tel,lower} = 0.8 f_{fib} + 4$ and $\rm f_{tel,upper} = 2 f_{fib} + 3$. For the smallest $50~\upmu$m fibers $\rm f_{tel,lower} = 1.5 f_{fib} + 4$, with no upper limit. The case for the $100~\upmu$m fiber core is intermediate to these two limiting cases.

For a spectrograph fed directly by fibers injected with $\rm 3 < f_{fib} < 5$, we find that an f/8 telescope would support fiber diameters from 100 to 500 $\upmu$m within the optimized range. This enables a wide range of spatial and spectral resolution for survey optimization with the same telescope and spectrograph hardware. Finally we comment that while it is always preferable to reduce the number of optical elements when possible, the native telescope focal ratio can often be reimaged by a suitable focal reducer or expander. On the basis of our analysis we suggest such additional optics should be designed, for most applications, to deliver a focal surface feeding micro-lens arrays at f/8. MLA designs with additional, or aspheric surfaces may alter these conclusions, but such augmentation incurs additional cost and complexity in design, fabrication and assembly.

\section* {Acknowledgments} % equivalent to \section*{ACKNOWLEDGMENTS}
This research was supported by funds from the University of Wisconsin-Madison Graduate School, the U.S. National Science Foundation grants AST-1517006 and AST-1814682, and the South African National Research Foundation SARChI program.

\appendix

\section{Derivation of model parameters}
\label{app:derivation}

Here we calculate the relationships between the critical MLA parameters (surface curvature, thickness and spacing) as a function of the telescope input and fiber input focal-ratios and the size of the micro-image, which again we take to be just slightly under-sized with respect to the fiber core diameter. The fiber input f-ratio $\rm f_{\rm fib}$ is defined in the air while $\rm D_p$ and $\rm d$ defines the f-ratio in glass. Hence $\rm f_{\rm fib}$ is usually multiplied with refractive index of glass ($\rm n_g$) at the system wavelength. 

\subsection{Optical diameters}
\label{sec:diam}

Relevant diameters can be computed by using the conservation of entropy (grasp) to relate the bi-convex MLA clear aperture $\rm d_a$ then can be formulated as a function of micro-image diameter $\rm d_m$:
\begin{equation}
\label{eq:da}
 \rm d_a \ = \ \frac{\rm f_{\rm tel}}{(\frac{\rm f_{\rm fib}}{\rm d_m})} \ = \rm d_m \ \upeta.
\end{equation}
From this, the beam diameter for a single field point at the input of the Plano-convex MLA (PC) is $\rm d \leq d_a - d_m$. This follows by construction in Figure~\ref{fig:mla}  due to the constraints that (i) the PC MLA diameter cannot not exceed that of the BC MLA  diameter, and (ii) the telecentricity requirement stipulates the principle ray for the edge field-point must fall at a height of $d_m$/2 from the optical axis while the distance between the marginal and principle ray is d/2. Hence d/2 + $d_m$/2 has to be equal to $d_a$/2, namely the clear aperture diameter of the PC and BC lenslets. 
Although $\rm d$ can be less than $\rm d_a - d_m$, this reduces $\rm r_p$. Hence maximizing $\rm d$ helps minimize SA, so we take 
$\rm d = d_a - d_m$. This condition for $\rm d$ can be rewritten with equation~\ref{eq:da} as
\begin{equation}
\label{eq:d}
 \rm d \ =  \rm d_m \  (\upeta - 1).
\end{equation}

\subsection{PC MLA thickness and radius of curvature}
\label{sec:pcmla}

From fiber input focal ratio, one can deduce that the plano-convex lens thickness is simply:

\begin{equation}
  \rm  f_{\rm fib} \ n_g = \frac{\rm D_p}{d}.  
\end{equation}

Note that since f-ratio is determined in air, refractive index has to be multiplied to get the f-ratio in the glass. This can be rewritten using equation~\ref{eq:d} as:

\begin{equation}
\label{eq:dp}
 \rm D_p \ = \rm d_m \rm n_g \rm f_{\rm fib} \ (\upeta - 1).
\end{equation}

Using the lens makers equation for plano convex lens with $\rm r_p$ is the radius curvature of the curved surface and the other side is flat, we can state that,

\begin{equation}
    \frac{1}{\rm D_p / \rm n_g} = (\rm n_g - 1)(\frac{1}{r_p} - \frac{1}{\infty}).
\end{equation}

Again note the use of $\rm n_g$ as a dividing factor as the focal length is usually defined in air but in our case the focus is the flat back plane of the MLA. Hence the thickness of PC lenslets should in air focal length multiplied by the refractive index of the glass. Hence, the plano-convex MLA (PC) radius of curvature can then be written as:

\begin{equation}
 \rm r_p \ = \rm D_p \rm (n_g -1)/ \rm n_g \ = \rm {d_m} \rm f_{\rm fib} \rm (n_g - 1) (\upeta - 1). 
\label{eq:rp}
\end{equation}

\subsection{MLA spacing}
\label{sec:spacing}

Again, using the lens makers equation for plano convex lens we can find that,

\begin{equation}
    \frac{1}{\rm D_G} = (\rm n_g - 1)(\frac{1}{r_p} - \frac{1}{\infty}).
\end{equation}

Since $\rm D_G$ is defined in air we don't need to include $\rm n_g$. Hence the gap between the bi-convex and plano-convex MLA can be derived as
\begin{equation}
 \rm D_G \ = \rm r_p/ \rm(n_g -1) \ = \rm {d_m} \rm f_{\rm fib} \ (\upeta - 1).
 \label{eq:dg}
\end{equation}

\subsection{BC MLA thickness and radius of curvature}
\label{sec:bcmla}

Similar to plano-convex MLA, the bi-convex MLA (BC) thickness can be derived as:
\begin{equation}
 \rm D_b \ = \rm n_g \rm d \rm f_{\rm tel} \ = \rm d_m \rm n_g \rm f_{\rm tel}  \ (\upeta - 1).
 \label{eq:db}
\end{equation}

\begin{figure}[H]
\centering
\includegraphics[width = 0.8\linewidth]{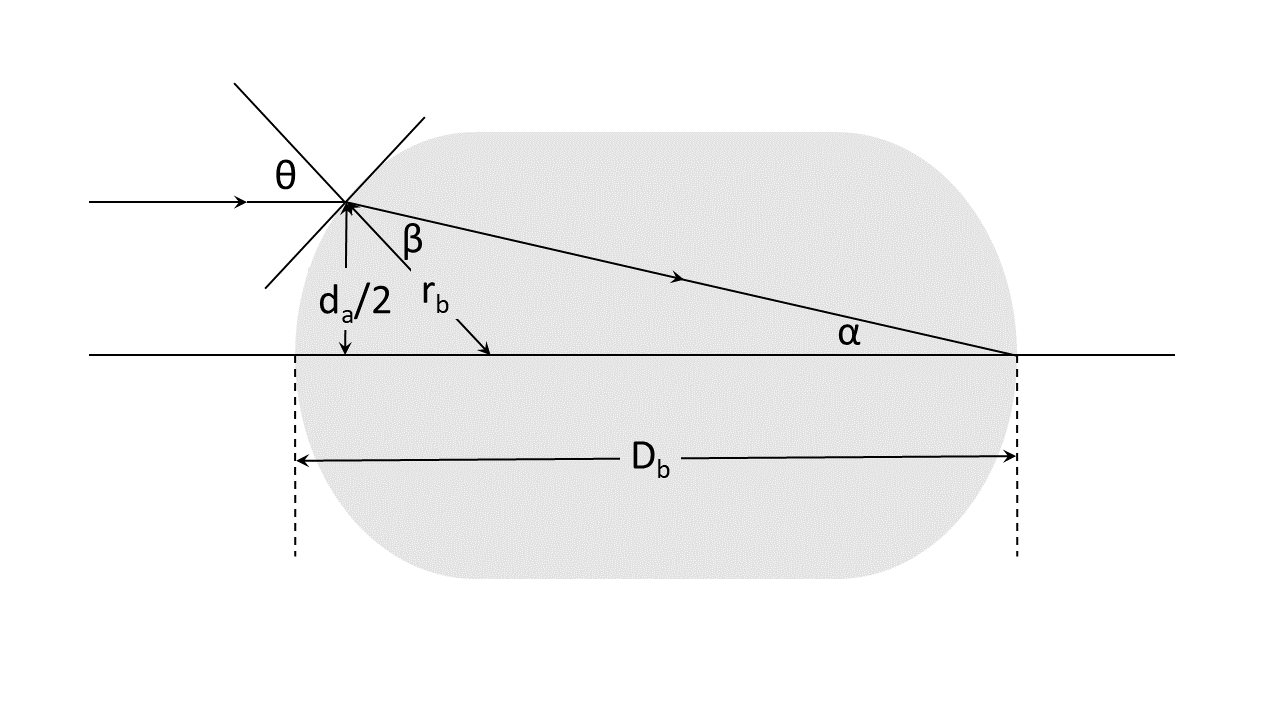}
\caption{Propagation path of edge field chief ray inside the bi-convex lenslet.}
\label{fig:bcmla}
\end{figure}

The bi-convex MLA (BC) radius of curvature can be derived from the path of chief ray of the edge field inside the BC MLA as shown in Figure~\ref{fig:bcmla}, as follows. From Snell's law, $\rm \sin{\uptheta} = n_g \sin{\upbeta}$, and from Figure~\ref{fig:bcmla} we have $\rm \uptheta = \upalpha+\upbeta$.

From this, we can easily derive
\begin{equation}
 \tan{\upbeta} = \frac{\sin{\upalpha}}{n_g - \cos{\upalpha}},
 \label{eq:tbeta1}
\end{equation}
while from the right angle triangle we can define $\rm \sin{\upalpha}$ as 
\begin{equation}
 \sin{\upalpha} \approx \frac{\rm d_a/2}{\sqrt{\rm (d_a/2)^2 + \rm D_b^2}},
  \label{eq:sbeta}
\end{equation}

%and similarly,

%\begin{equation}
% \cos{\upalpha} \approx \frac{\rm D_b}{\sqrt{\rm (d_a/2)^2 + \rm D_b^2}}.
%\end{equation}

and similarly,

\begin{equation}
 \sin(\upalpha+\upbeta) = \frac{\rm d_a/2}{\rm r_b},
 \label{eq:sba}
\end{equation}

which we can combine to write:

\begin{equation}
 \tan{\upbeta} = \frac{\rm d_a/2}{\sqrt{(\rm n_g \rm r_b)^2 - (\rm d_a/2)^2}}.
  \label{eq:tbeta2}
\end{equation}
From equations \ref{eq:tbeta2} and \ref{eq:tbeta1} we can easily define the radius of the BC MLA as,
\begin{equation}
 \rm r_b \ = \ \frac{1}{\rm n_g} \left[\left(\rm n_g \sqrt{\rm (d_a/2)^2 + \rm {D_b}^2} - \rm D_b\right)^2 + \rm (d_a/2)^2\right]^{1/2}
\label{eq:rb1}
\end{equation}
for which equations~\ref{eq:da} and \ref{eq:db} can be used to express $\rm r_b$ in terms of $\rm f_{\rm tel}$, $\rm f_{\rm fib}$ (or $\rm f_{\rm tel}$ and $\rm \upeta$), and $\rm d_m$:
\begin{equation}
 \rm r_b \ = \rm d_m \ \upeta \left[\left(\sqrt{1/4+(\rm n_g \ \rm f_{\rm fib} \ (\upeta-1))^2} - \rm f_{\rm fib} \ (\upeta-1)\right)^2+1/(4{\rm n_g^2})\right]^{1/2}. 
\label{eq:rb2}
\end{equation}

\section{Refraction on the spherical surface: Dependence of microlens thickness on beam diameter.}
\label{ap:focal length}

\begin{figure}[h]
\centering
\includegraphics[width = 0.6\linewidth]{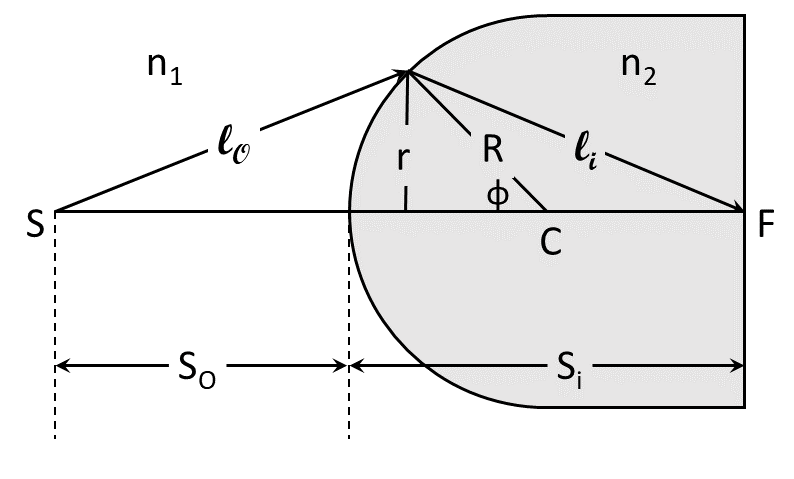}
\caption{Derivation of plano-convex microlens focal length.}
\label{fig:fl}
\end{figure}

Here we define the PC MLA thickness, f (sag plus flat thickness), with radius of curvature R, such that the ray of an object at infinity at height r crosses the optical axis at the exit surface. Due to SA, this thickness depends on height. We have defined the thickness for an object at infinity and in line with our requirement of treating the PC MLA which is the imaging lens in our system. This generic description of MLA thickness will help us define the PC lenslet parameters requried to locate the CoC on the PC MLA exit surface in the next subsection.
Using the derivation from Hecht\cite{hecht} for an object at a distance $\rm S_o$ in front of the MLA, illustrated in Figure~\ref{fig:fl}, 
their equation 5.4, given here as:
\begin{equation}
\label{eq:hecht}
   \frac{\rm n_1 (\rm S_o + \rm R)}{\rm l_o} = \frac{\rm n_2 (\rm S_i - \rm R)}{\rm l_i}.
\end{equation}
% For a collimated input beam, $\rm s_o = l_o = \infty$ and $\rm s_i = f$. For $\rm n_1$ = 1 and $\rm n_2 = n_g$, the equation % %\ref{eq:hecht} can be written as,
can be rewritten for an object at infinity ($\rm S_o = l_o = \infty$, $\rm S_i = f$, $\rm n_1$ = 1 and $\rm n_2 = n_g$):
\begin{equation}
 \rm n_g (\rm f - \rm R) = \rm l_i. 
\end{equation}
It is straight forward to construct:
\begin{equation}
 \rm l_i = [\rm R^2 + (\rm f - \rm R)^2 + 2 \rm R (\rm f - \rm R)^2 cos\upphi]^{1/2},
\end{equation}
where $\rm \cos\phi = \sqrt{1 - (\frac{r}{R})^2}$. Solving for f - R we arrive at an expression for f:
\begin{equation}
 {\rm f} = {\rm R} \left[1+\frac{1}{\sqrt{\rm n_g^2 - (\frac{\rm r}{\rm R})^2} - \sqrt{1 - (\frac{\rm r}{\rm R})^2}} \right]
\end{equation}
which has the expected paraxial approximation (r/R$\ll$1) of $\rm f = R (n_g / (n_g-1)$. The dependence of the microlens thickness f on (r, R, $\rm n_g$) is simply the manifestation of SA on a collimated beam. 

\section{Effect of spherical aberration on microlens focal length: Longitudinal and Transverse spherical aberration}
\label{ap:TSALSA}

The longitudinal SA (LSA) is defined by the distance between paraxial focus and the focus of the marginal ray while the transverse SA (TSA) is the marginal ray height at the paraxial focus. However, to minimize the effect of SA the location and radius of circle of least confusion (CoC) is important to understand. The location would directly define the thickness of the PC lenslet while the CoC radius would impose condition on the lenslet properties in order to restrict the micro-image to form within the fiber core to reduce photon loss at the MLA-fiber junction. Increasing the fiber core diameter arbitrarily to engulf entire micro-image would lead to increase in optical entropy which can be equated to loss of observation time. 

\begin{figure}[h]
\centering
\includegraphics[width = 0.8\linewidth]{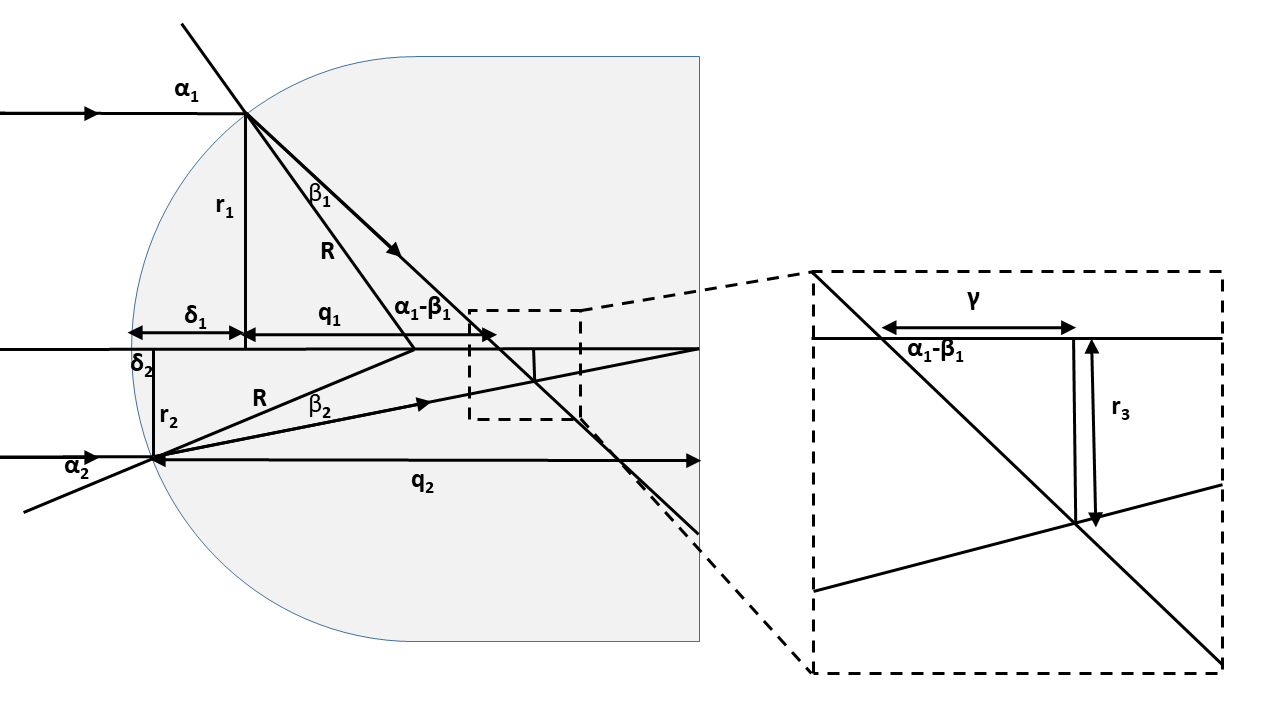}
\caption{Derivation of position and radius of circle of least confusion.}
\label{fig:sa}
\end{figure}

Now having the analytical description of thickness of PC lenslet in our arsenal, we will describe the radius and position of circle of least confusion and what should be the strategy for minimizing the radius. In figure \ref{fig:sa} we define a plano-convex microlens of radius of curvature R fed with a collimated beam. The height of the ray from the optical axis is r; $\rm \upalpha$ and $\rm \upbeta$ are the angles of incidence and refraction at the input surface, respectively; q and $\rm \updelta$ are the distances of the crossing of optical axis inside the microlens and chief ray entry point to the arbitrary ray entry point respectively. Suffixes 1 and 2 denote angles and dimensions associated with marginal and paraxial ray, respectively. The radius of the circle of least confusion (CoC) is defined as $\rm r_3$ while the distance from circle of least confusion from the marginal ray focus is $\rm \upgamma$.

It is easy to find that $\rm \updelta_2 = R - \sqrt{R^2 - r_2^2}$. We can also deduce the following,
\begin{equation}
 \frac{\rm r_3}{\upgamma} = \frac{\rm r_1}{\sqrt{\rm R^2-\rm r_1^2}},
\end{equation}
and,
\begin{equation}
 \frac{\rm r_2}{\rm r_3} = \frac{\rm f(\rm r_2) - \updelta_2}{\rm f(\rm r_2) - \rm f(\rm r_1) - \upgamma}.
\end{equation}
Solving equation 17 and 18 we get,
\begin{equation}
 \rm r_3 = \frac{\rm r_1 \rm r_2 \left(\rm f(\rm r_2) - \rm f(\rm r_1)\right)}{\rm r_1\left[\rm f(\rm r_2) - \rm R + \sqrt{\rm R^2 - \rm r_2^2} \right] + \rm r_2\sqrt{\rm R^2 - \rm r_1^2}},
\label{eq:r3}
\end{equation}
and,
\begin{equation}
 \upgamma = \frac{\rm r_2\sqrt{\rm R^2 - \rm r_1^2}(\rm f(\rm r_2) - \rm f(\rm r_1))}{\rm r_1 \rm f(\rm r_2) - \rm r_1(\rm R-\sqrt{\rm R^2 - \rm r_2^2}) + \rm r_2(\rm R-\sqrt{\rm R^2 - \rm r_1^2})}
\label{eq:gamma}
\end{equation}

One can differentiate equation \ref{eq:r3} and find the maxima to identify the location of the CoC. However, this analytical approach is quite cumbersome and hence it is better to solve the problem numerically.

\pagebreak

% References
\bibliography{report} % bibliography data in report.bib
\bibliographystyle{spiejour}

\end{document}